\documentclass[journal]{IEEEtran}
\usepackage{lipsum}

\ifCLASSINFOpdf
   \usepackage[pdftex]{graphicx}

\else
\fi

\usepackage{placeins}

\usepackage{amsmath}
\usepackage{multirow}

\ifCLASSOPTIONcompsoc
  \usepackage[caption=false,font=normalsize,labelfont=sf,textfont=sf]{subfig}
\else
  \usepackage[caption=false,font=footnotesize]{subfig}
\fi

\usepackage{tikz}

\usepackage[latin1]{inputenc}

\usetikzlibrary{shapes,arrows}
\newcommand{\norm}[1]{\left\lVert#1\right\rVert}

\begin{document}

\title{Certainty based reduced sparse solution for dense array EEG source localization}

\author{Teja Mannepalli, Aurobinda Routray}%
%

\maketitle

\begin{abstract}
	The EEG source localization is an ill-posed problem. It involves estimation of the sources which outnumbers the number of measurements. For a given measurement at given time all sources are not active which makes the problem as sparse inversion problem. This paper presents a new approach for dense array EEG source localization. This paper aims at reducing the solution space to only most certain sources and thereby reducing the problem of ill-posedness. This employs a two-stage method where the first stage finds the most certain sources that are likely to produce the observed EEG by using a statistical measure of sources, the second stage solves the inverse problem by restricting the solution space to only most certain sources and their neighbors. This reduces the solution space for other source localization methods hence improvise their accuracy in localizing the active neurological sources in the brain. This method has been validated and applied to real 256 channel data and the results were analyzed. 
\end{abstract}


\IEEEpeerreviewmaketitle

\section{Introduction}

\IEEEPARstart{T}{he} study of active neurological sources in the brain is essential to know about various physiological, cognitive aspects and also brain abnormalities such as focus, seizures, tumors and epilepsy\cite{sanei2013eeg} \cite{nolte2002human}. The estimation of these active sources is termed as source localization. This involves solving the forward problem and also the inverse problem. The forward problem is finding the lead field matrix that represents the behavior of electrical activity in the sources. There are two common ways of solving this, dipole fitting model or distributed source model. In dipole fitting models\cite{sommariva2014sequential}, one or few number of dipoles are assumed to represent the sources, their orientation, location, and magnitudes are found out. This is formulated as a non-linear least squares problem which requires prior estimation of the number of sources. In distributed source model  \cite{michel2004eeg} \cite{pascual1994low}, the brain is 3D tessellated into voxels (small cubes) and each voxel is considered as having a source with three dipole moments along x, y and z (or only one dipole moment \cite{phillips2002anatomically}). The problem reduces to a linear one with only magnitudes of the dipole moment(s) of sources as unknown. 

The inverse problem involves finding the sources that are responsible for the generation of EEG. This whole problem can be formulated as a linear, instantaneous model:

\begin{equation}
\mathbf{\Phi = KJ + n}
\end{equation}

$\mathbf{\Phi}$ is the measurement vector of size $N \times 1 $ at a time instant, $N$ is the number of electrodes, $\mathbf{K}$ is the lead field matrix of size $N \times 3M$ generated by solving the forward problem, $M$ is the number of grids or voxels (each grid is assumed having a source of dipole moment either along x,y and z making it $3M$ or only perpendicular to cortical sheet, thus $M$\cite{phillips2002anatomically}), $\mathbf{J}$ is the unknown current density vector of size $3M \times 1$ (or$M\times 1$), $\mathbf{n}$ is some unknown noise. 

The essence of inverse problem is finding the unknown $\mathbf{J}$ with known $\mathbf{K}$ and given $\Phi$. One way of solving the inverse problem is:
\begin{equation}
\mathbf{F} = \underset{\mathbf{J}}{\text{min}}\mathbf{J^TWJ}\hspace{5pt} \text{such that}\hspace{5pt} \mathbf{\Phi=KJ}
\label{CMP}
\end{equation}

$\mathbf{F}$ is the cost function. This is a minimization problem under the equality constraint. The solution for this is:

\begin{equation}
\mathbf{\hat{J}=T\Phi \hspace{5pt} \text{where} \hspace{5pt} T = W^{-1}K^T{[KW^{-1}K^T]}^+}
\label{ESoL}
\end{equation}.

The minimum norm estimate (MNE) solves the above by assigning $\mathbf{W}$ as identity matrix. The weighted minimum norm estimate(WMNE) solves by weighing $\mathbf{ W }= diag(\mathbf{||K_i||^2}) $ , $\mathbf{ K_i} $ is the $i_{th}$ column of $\mathbf{K}$ to compensate the deep sources.
	
Low Resolution impedance Tomography (LORETA) \cite{pascual1994low} solves by assigning $ \mathbf{W = B}diag(\mathbf{||K_i||})) $, $\mathbf{B}$ is discrete spatial laplacian operator. This $\mathbf{B}$ incorporates the assumption that each source is synchronized with it's neighboring one. 
	
Focal under-determined system solver (FOCUSS) \cite{gorodnitsky1997sparse},\cite{gorodnitsky1995neuromagnetic} is a powerful and intelligent method that works on iterative convergence of the solution by itself aiming a sparse one. However, this algorithm is unstable in the presence of noise and may provide a scattered and non-clustered solution missing the gradient in it thus making hard to interpret.

\begin{align*}
\mathbf{C_i} = diag & (\mathbf{J_{i-1}(1),\dots,J_{i-1}(3M)}) \\
& \mathbf{q_i} = (\mathbf{KC_i})^+\mathbf{\Phi} \\
& \mathbf{J_i} = \mathbf{C_iq_i}
\label{FOCUSS}
\end{align*}

$ \mathbf{C_i} $ is weight based on the previously obtained solution of $\mathbf{J}$.
Eq. (3) can also be presented as an unconstrained minimization problem:
\begin{equation}
\mathbf{F = \underset{J}{\text{min}}{||\Phi - KJ||^2_a} + \alpha\mathbf{||J||^p_b}}\end{equation}
where $\alpha$ is the regularization parameter. $||\Phi - KJ||^2_a$ is loss function and $\alpha\mathbf{||J||^p_b}$ is regularization term.

standardized low resolution impedance tomography (sLORETA)\cite{pascual2002standardized} uses euclidean norm ($l_2$) on both loss function and regularization (zero-order tikhonov-philips regularization). \cite{phillips2002systematic}.
The exact inverse solution is given by:
\begin{equation}
\mathbf{\hat{J} = K^T[KK^T}+\alpha\mathbf{ I]}^{-1}\mathbf{\Phi}
\end{equation}
This can explain the measurement vector under the presence of noise with gradient due to $l_2$ norm regularization in it despite producing a non-sparse and blurred solution. The $l_2$ norm works well even when the data is highly correlated but is much sensitive to outliers. On the other hand, the $l_1$ norm \cite{tibshirani2005sparsity} is less sensitive to outliers hence can be more robust to noise \cite{silva2004evaluation}. The $l_1$ norm also provides a sparse solution which $l_2$ norm cannot. However, $l_1$ norm may not provide a good estimate when the data is highly correlated. Due to all this, the mixed norms ($l_{12}$ or $l_{21}$) can be used \cite{gramfort2012mixed} \cite{gramfort2013time} \cite{nie2010efficient} \cite{strohmeier2014improved} on regularizing the solution and also on loss function to utilize the advantages provided by both $l_2$ and $l_1$ norms. The $l_{12}$ and $l_{21}$ norms can be applied only on matrices, so $\mathbf{\Phi}$ is considered $N\times T$ instead of $N\times 1$ to extract the block row structure $l_{21}$ norm can provide\cite{gramfort2013time}. $T$ is the time course of EEG. $l_p$ norm sparse solution \cite{xu2007lp} in uses $l_p(p\leq 1)$ norm as the constraint to regularize which also provides a sparse solution like $l_1$ norm.

This inverse problem can be formulated in bayesian theorem for probability \cite{baillet2001electromagnetic}. The anatomical and physiological constraints can be incorporated into the prior distribution of the source. This finds the probability of the sources being active rather than finding the active ones.

The EEG data can be considered a stationary signal for some $ms$ of time depending on the kind of experiment being conducted \cite{sanei2013eeg}. dMAP-EM \cite{lamus2012spatiotemporal}, STOUT \cite{castano2015solving}, \cite{sorrentino2013dynamic} (particle filter), \cite{long2011state} (kalman filter) take advantage of this temporal coherence in the EEG channels to solve the problem. In \cite{costa2017bayesian}, a structured sparsity prior\cite{zhang2011sparse} is used to incorporate the sparse nature of every source. In \cite{costa2015sparse} an $l_0 + l_1$ norm is used to regularize the solution space and bernouli-laplacian prior is employed to incorporate sparse nature of the sources. In \cite{gavit2001multiresolution}, the source space's resolution is iterative updated using the regularized solution calculated in every iteration for that space. In \cite{friston2008hierarchical} a hierarchal modeling in the brain is employed. The EEG data can be considered as a tensor (an $n$ dimensional array) by considering channels, time and frequency as three dimensions.\cite{becker2014eeg} \cite{cong2015tensor}.

The ill-posedness ($3M\gg N$) of this problem is making it difficult to find the active sources accurately as many number of solutions will be possible. The brain is tessellated as a $M$ voxels using a 3D grid. Each voxel is modeled by a source with dipole moment of some magnitude and orientation. A fixed location is modeled as a source either with three dipoles moments along x,y and z or only one dipole with moment along z that is perpendicular to the cortical surface instead of three\cite{phillips2002anatomically} thereby reducing the number of unknowns. Also, as the system is under determined, the solution space has to be constrained based on anatomical and physiological aspects of the brain \cite{phillips2002anatomically} \cite{phillips2002systematic}. More review on EEG source localization algorithms is presented in \cite{grech2008review}, \cite{baillet2001electromagnetic}, \cite{awan2018recent}. 

The method presented here utilizes sparse nature of the active sources which makes each of the source's signature in $\mathbf{\Phi}$ evident and aims at reducing the solution space only those most certain sources that might be active.

\begin{figure*}[!t]
	\centering
	\subfloat{(a)}{%
		\includegraphics[width=2.2in]{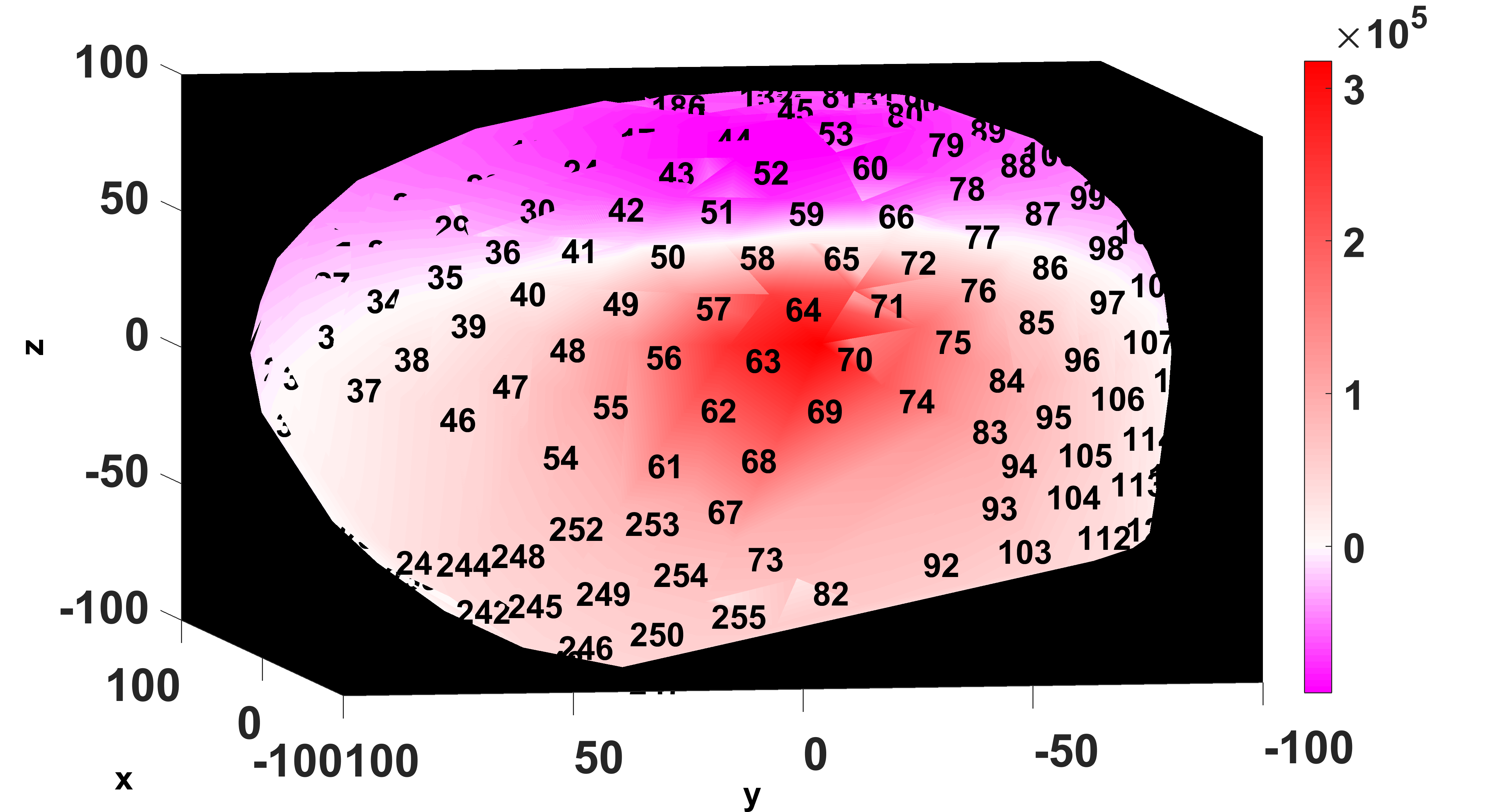}%
		\label{M11}}\hfill
	\subfloat{(b)}{%
		\includegraphics[width=2.2in]{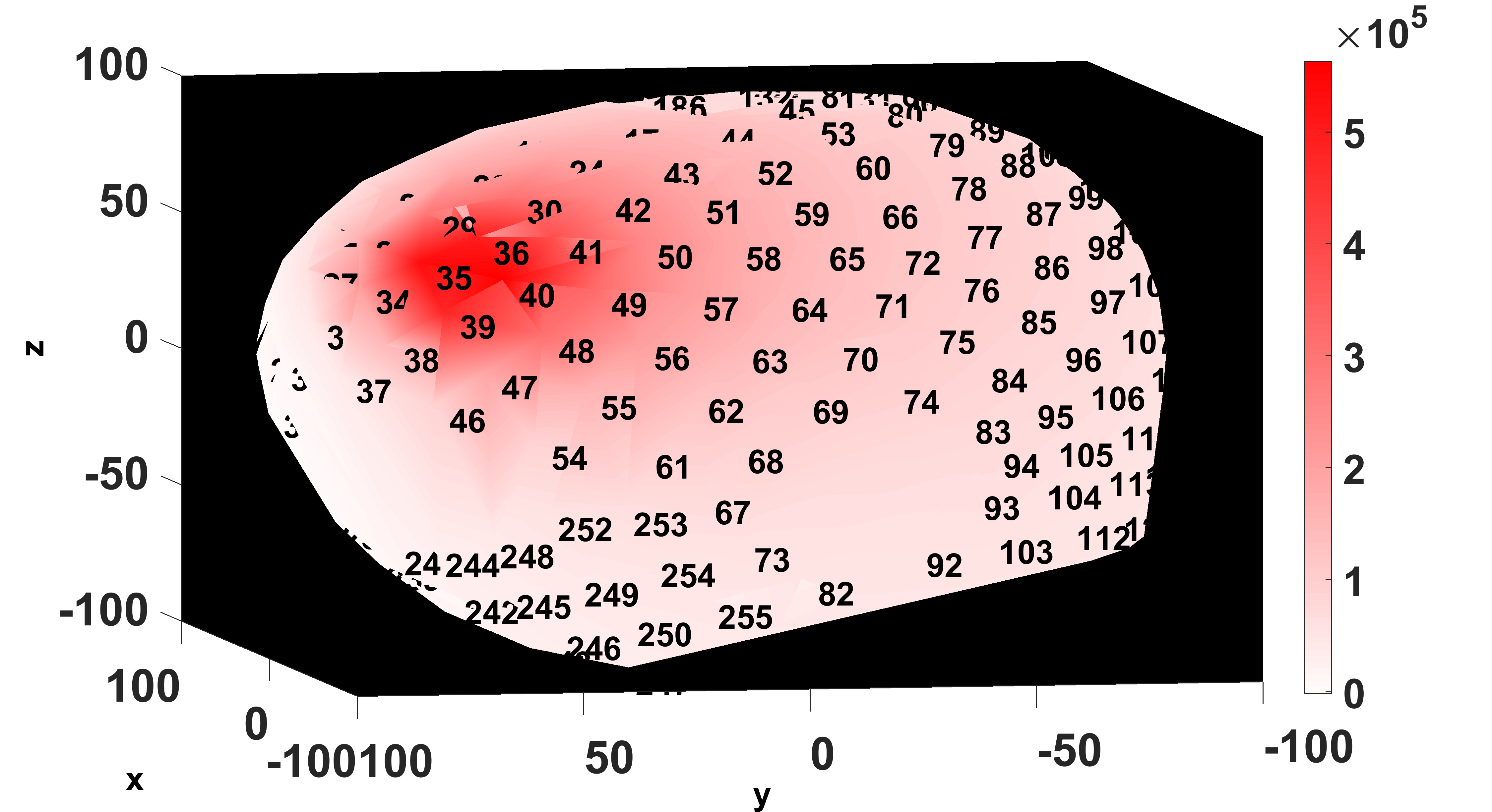}%
		\label{M12}}\hfill
	\subfloat{(c)}{%
		\includegraphics[width=2.2in]{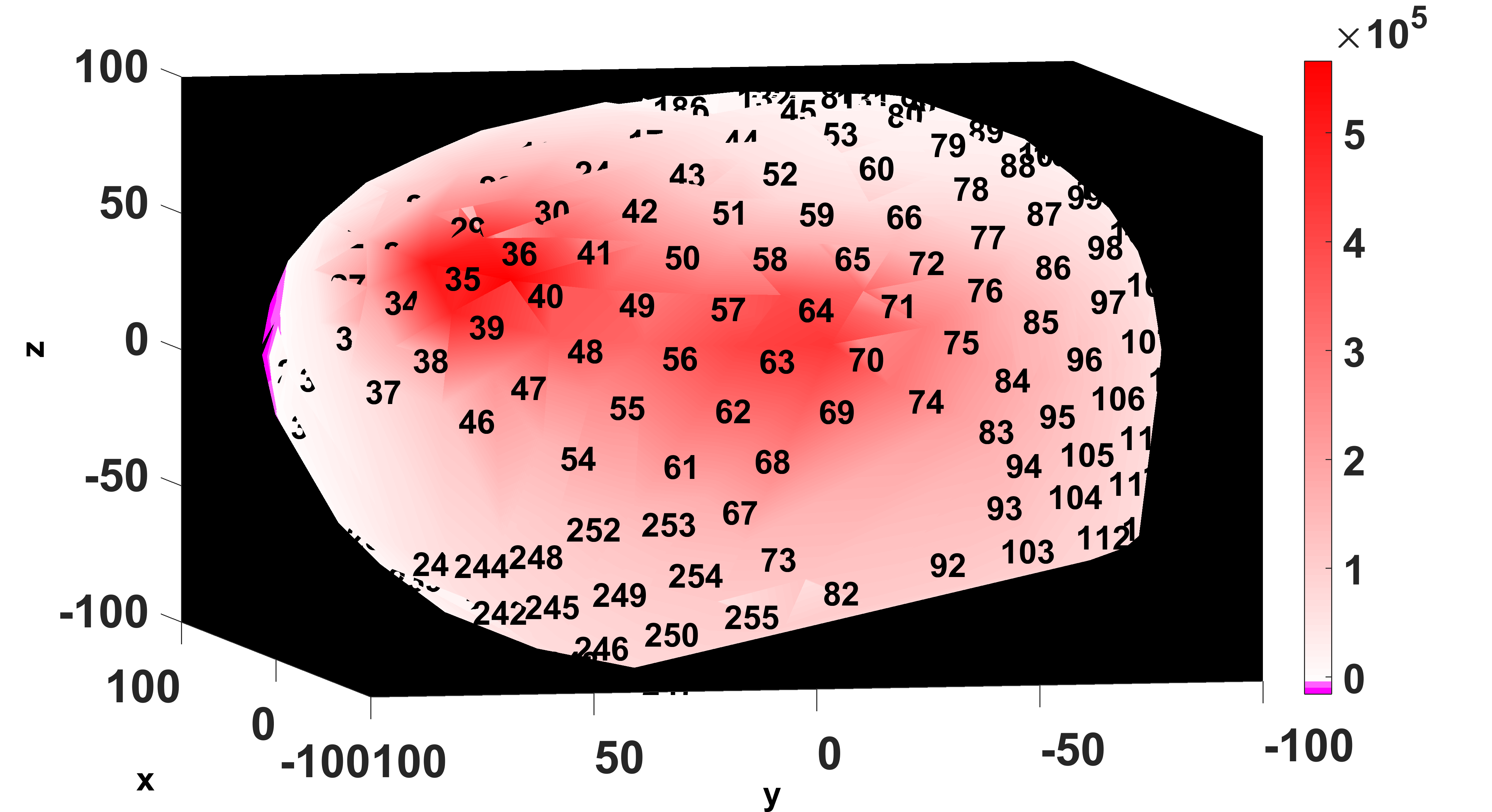}%
		\label{M13}}\hfill
	\centering
	\caption{The potential distribution(source signature) of the sources located at  (a) [-35.4 53.8 31.6] with moment [0,1,0] and (b) [-57.7 -1.9 31.6] with moment [0,0,1]. The source signature of simultaneous activation of both is shown in (c) where the source signatures are distinctly separate.}
	\label{M1}
\end{figure*}

\section{Methodology}

The potential distribution of any source has only one prominent global maximum or minimum three-dimensionally (referred to as 'peak') on the scalp. As the sources which might be active will be few, every peak of the active sources will be prominently evident in $\mathbf{\Phi}$. This can be observed from Fig.\ref{M1}. Two sources located at [-35.4 53.8 31.6] mm (this source's distribution is peaked at $70^{th}$ electrode) and [-57.7 -1.9 31.6] mm (peaked at $35^{th}$ electrode) are excited separately, This is shown in \ref{M1}(a) and \ref{M1}(b). The simultaneous activation of both the sources is shown in Fig. \ref{M1}(c) where the source distributions are distinctly evident and peaked at both $70^{th}$ and $35^{th}$ electrode. Form the information of this peaks $35^{th}$ and $70^{th}$ in $\mathbf{\Phi}$, the sources which has these $35^{th}$ and $70^{th}$ as their peaks and has the same/similar distributions as that in $\mathbf{\Phi}$ might be the most certain sources that are active.

The method presented here utilizes the information provided by peaks in $\mathbf{\Phi}$ to find the most certain sources that have generated the peaks observed in $\mathbf{\Phi}$. This method has a preliminary stage that has to be done only at the beginning, stage-I reduces the solution space to most certain sources and stage-II solves the inverse problem by restricting the solution space to those most certain sources.

\textbf{Stage-0}:This is the preliminary stage. This starts with finding the nearest neighbors(neighboring regions) of every electrode. The neighboring ones of $i^{th}$ electrode located at $r_i$ are $||r_i - r_j|| \le \mu $, where $j=1 \dots N$. The nearest neighbors are found out by setting $\mu$ such that every electrode has one level of surrounding electrodes. As the electrodes lie on spherical surface, geodesic distance between neighboring ones is calculated.

The peak index for all sources are found out. The electrode is called a 'peak index' if it has the maximum(or minimum) power in its neighboring region. A source will have only one peak index. It can be either global maximum or minimum. Every source will have an electrode as it's peak. A source will have only one peak, but $\mathbf{\Phi}$ can have many peaks as many sources might be active.

\textbf{Stage-I:} The peak indexes in $\mathbf{\Phi}$ are found out. If an $i^{th}$ electrode is a peak in $\mathbf{\Phi}$, the sources with peak index at the same $i^{th}$ electrode can be possible sources which might be active. The certainties of these possible sources are found by comparing the hat (The peak index's electrode and its neighboring electrodes are collectively termed as hat) of $s_n$ and $\mathbf{\Phi}$ using a statistical index (this measures the similarity between them). This index is defined as:
\begin{equation}
	SI(\mathbf{\Phi}|s_n) = \sum_{i=1}^{n_e} \frac{1}{\exp{\frac{1}{b} (\frac{|sc. K_h(s_n,i)|-|\phi_h(i)|}{|\phi_h(i)|})^{2}}} , b>0
	\label{11}
\end{equation}
for $s_n$, $n_e$ is the number of neighbors for an electrode lying in its neighboring region. scale factor, $sc = \frac{\phi_h}{K_h}$ at $i=1$, the peak index electrode. 
This also can be reformulated using an $l_2$ norm to make the index more robust to noise.
\begin{equation}
	SI(\mathbf{\Phi}|s_n) = \norm{\frac{|sc|.|K_h(s_n)|-|\Phi_h|}{|\Phi_h|}}_2
\end{equation}

The primary possible sources are the above ones with more certainty values and all possible sources with less certainties will be removed. 

Under activation of many sources, each source's distribution may be interfered with other, hence the neighbors of above determined most certain sources are also be considered as most certain sources. The secondary possible sources are neighbors (with one inter-point grid distance or two) of these more certain primary possible sources. The secondary possible sources are assigned the same values as their primary possible sources. If a source lies both in primary and secondary possible sources, the maximum value of certainty among them will be assigned to it. Only these primary and secondary possible sources are used in stage-II.

\textbf{Stage-II:}
The solution space is reduced. The cost function minimized is:
\begin{equation}
\underset{\mathbf{J_r}}{min}{||\mathbf{\Phi}-\mathbf{K_rJ_r}||^2}
\end{equation}
where $\mathbf{K_r}$ of size $N\times a$ is the lead field matrix restricts the solution space to only the more certain sources ($a$) found in stage-I.
The initial estimate to the above problem is $\mathbf{J_{prior}}$ which contains certainties of the prior sources. To solve the above cost function any source localization algorithm like FOCUSS, sLORETA etc can be used.

\section{Simulations and validation}

\subsection{Simulated data tests}
The head model is generated by three-concentric spherical head model\cite{salu1990improved}. The head model contains three layers brain, skull and scalp. Their radii are 80, 85 and 92 mm and their conductivities are 0.33, 0.0042, and 0.33 $\Omega^{-1} m^{-1}$. The electrodes are aligned to CTF coordinate system. The number of EEG channels are 256. The channel locations are based on 10-10 equivalent system for hydrocel geodesic sensor nets. The grid assumed is a regular 3D grid of inter grid-point distance of 11.2 mm. The total number of grid points are 1535 ($M$). The size of $\mathbf{K}$ is $256\times4605$ ($3M$). The lead field matrix is generated by three concentric spherical model. FOCUSS and sLORETA are implemented using \cite{gorodnitsky1997sparse} and \cite{pascual2002standardized} where $\alpha$ is found using cross validation error method. The initial estimate of FOCUSS is all the sources. This is because if FOCUSS is initialized with an incomplete solution which missed out a source, It is very unlikely to retrieve that source. The maximum number of iterations for FOCUSS is 100 and the iteration is stopped when the difference between successive errors $||J_{i}-J_{i-1}||^2\le 10^{-8}$. In Stage II, in absence of noise FOCUSS is used, while in presence of noise sLORETA is used.  

The evaluation indexes \cite{xu2007lp} used here are localization error ($E_{localization}$), source energy error ($E_{energy}$) and notional blurring index ($NBI$).

\begin{figure*}[!t]
	\centering
	\subfloat{(a)}{%
		\includegraphics[width=3.33in]{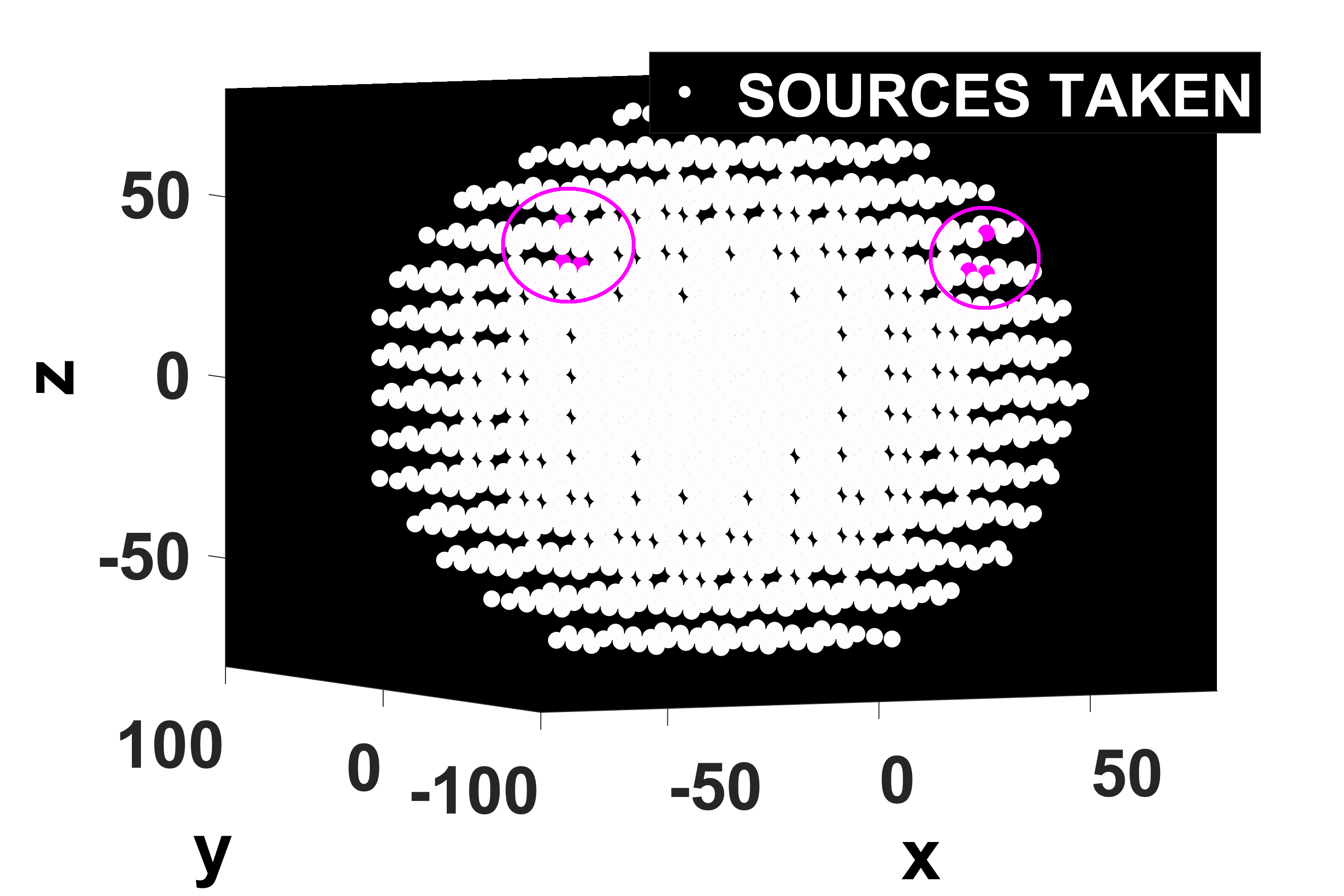}%
		\label{M41}}\hfill
	\subfloat{(b)}{%
		\includegraphics[width=3.33in]{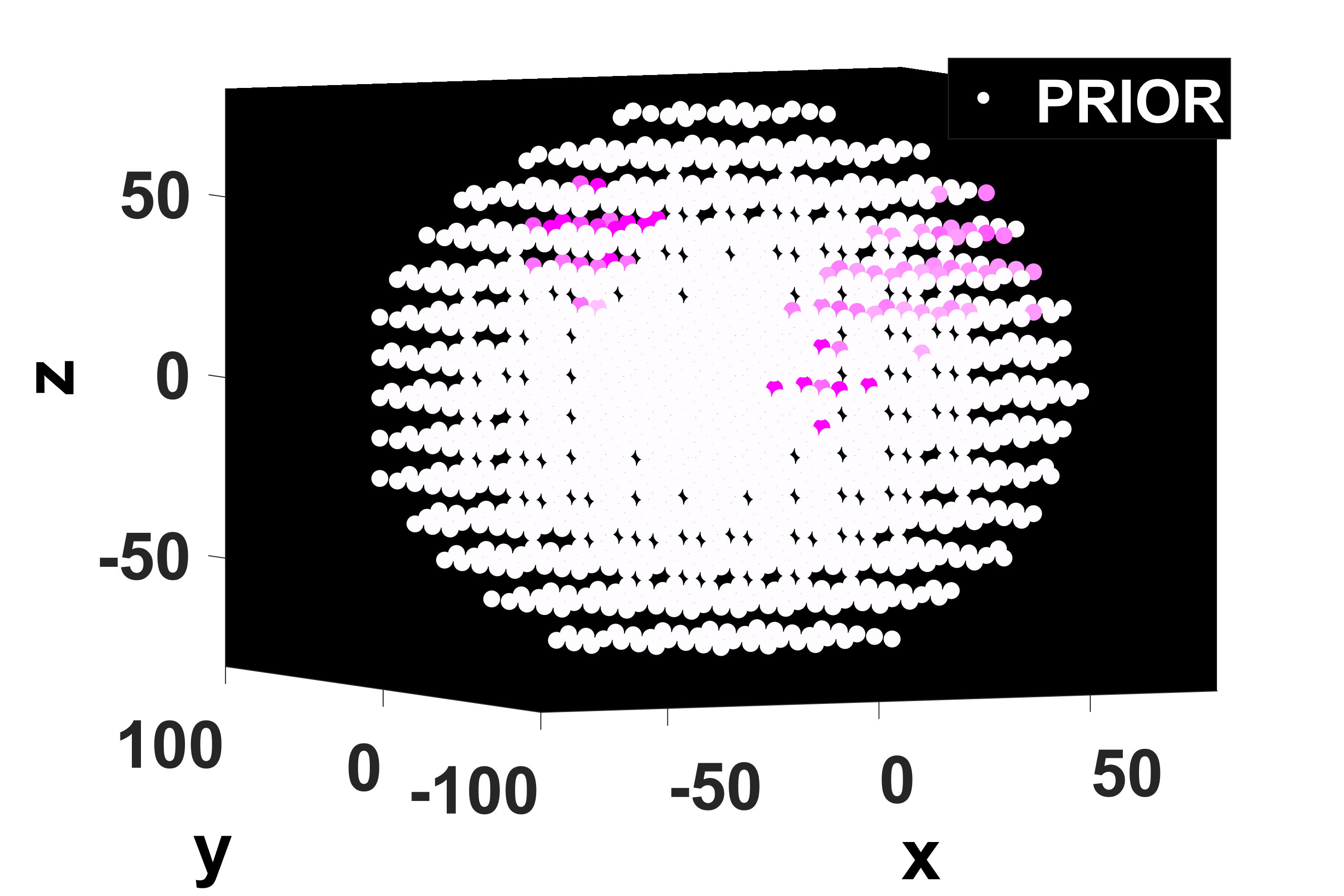}%
		\label{M42}}\hfill
	\subfloat{(c)}{%
		\includegraphics[width=3.33in]{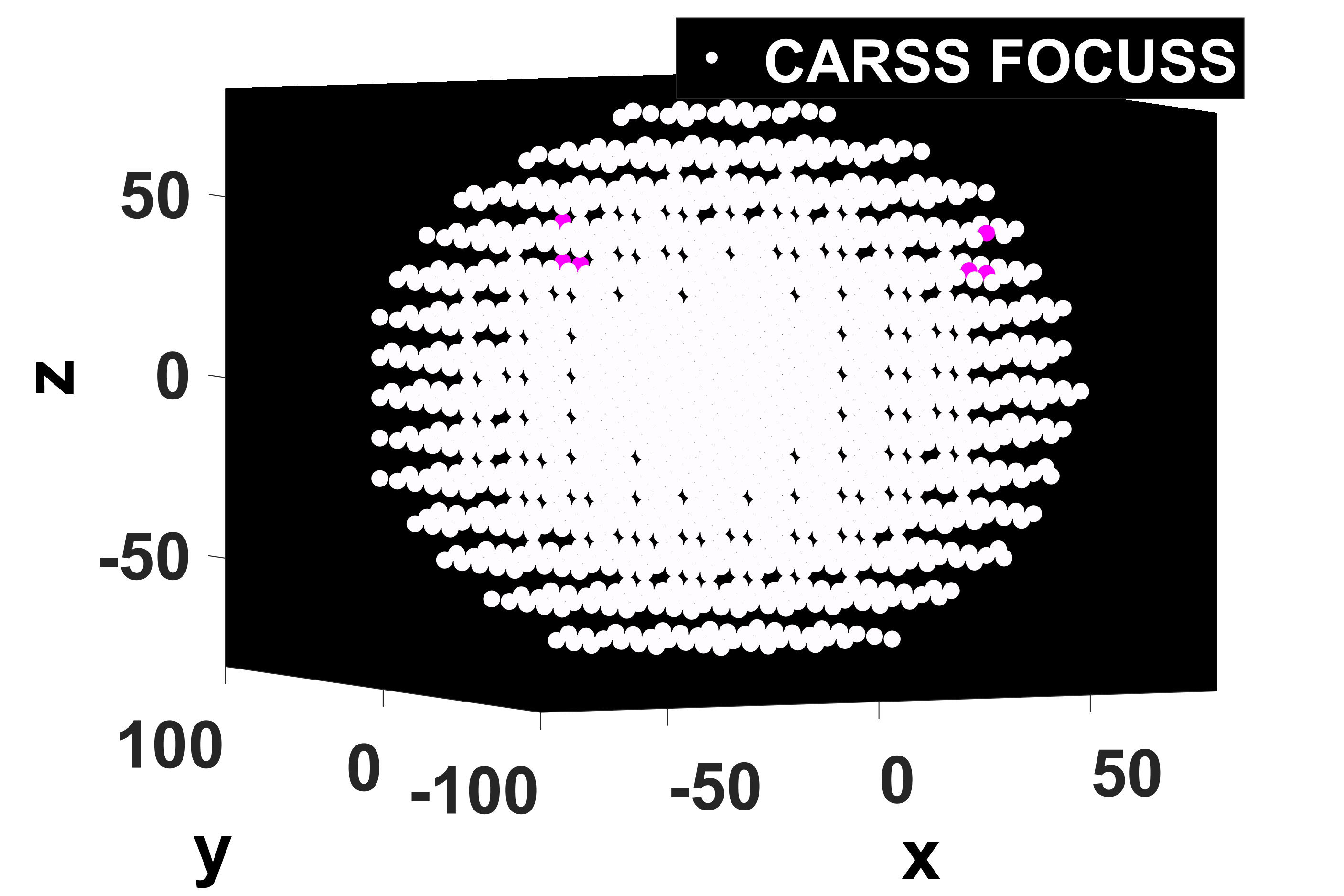}%
		\label{M43}}\hfill
	\subfloat{(d)}{%
		\includegraphics[width=3.33in]{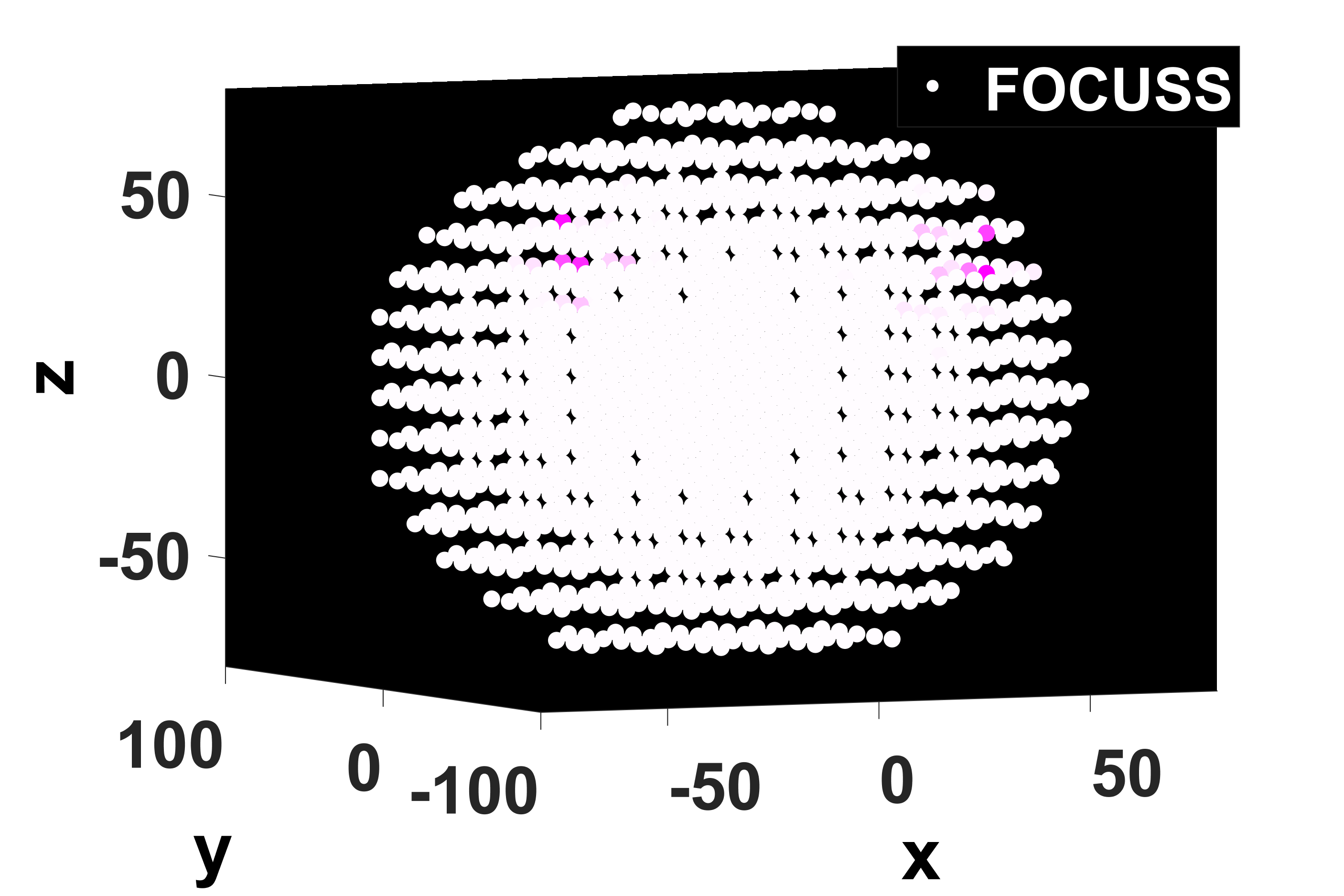}%
		\label{M44}}\hfill
	\subfloat{(e)}{%
		\includegraphics[width=3.33in]{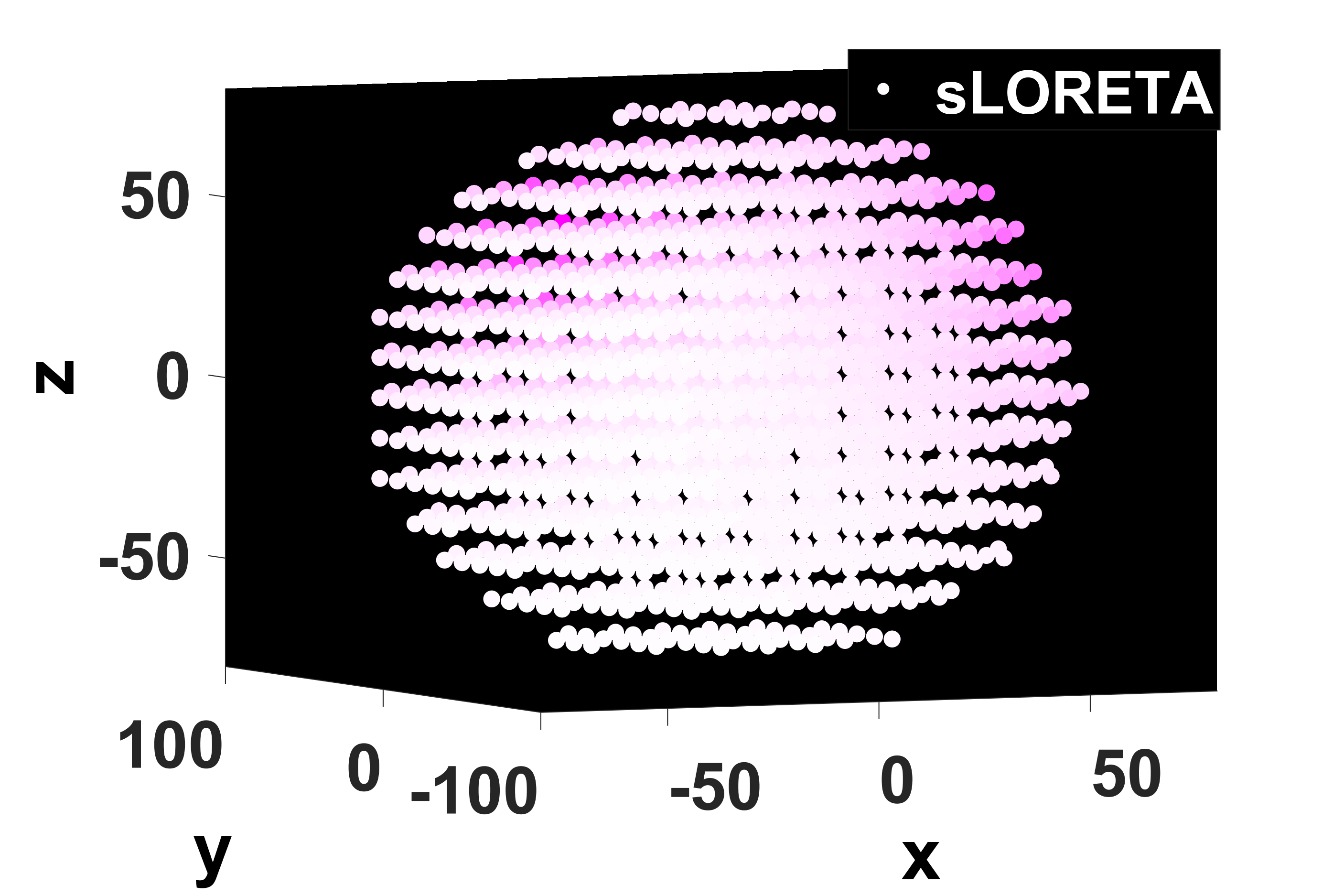}%
		\label{M45}}\hfill
	\subfloat{(f)}{%
		\includegraphics[width=3.33in]{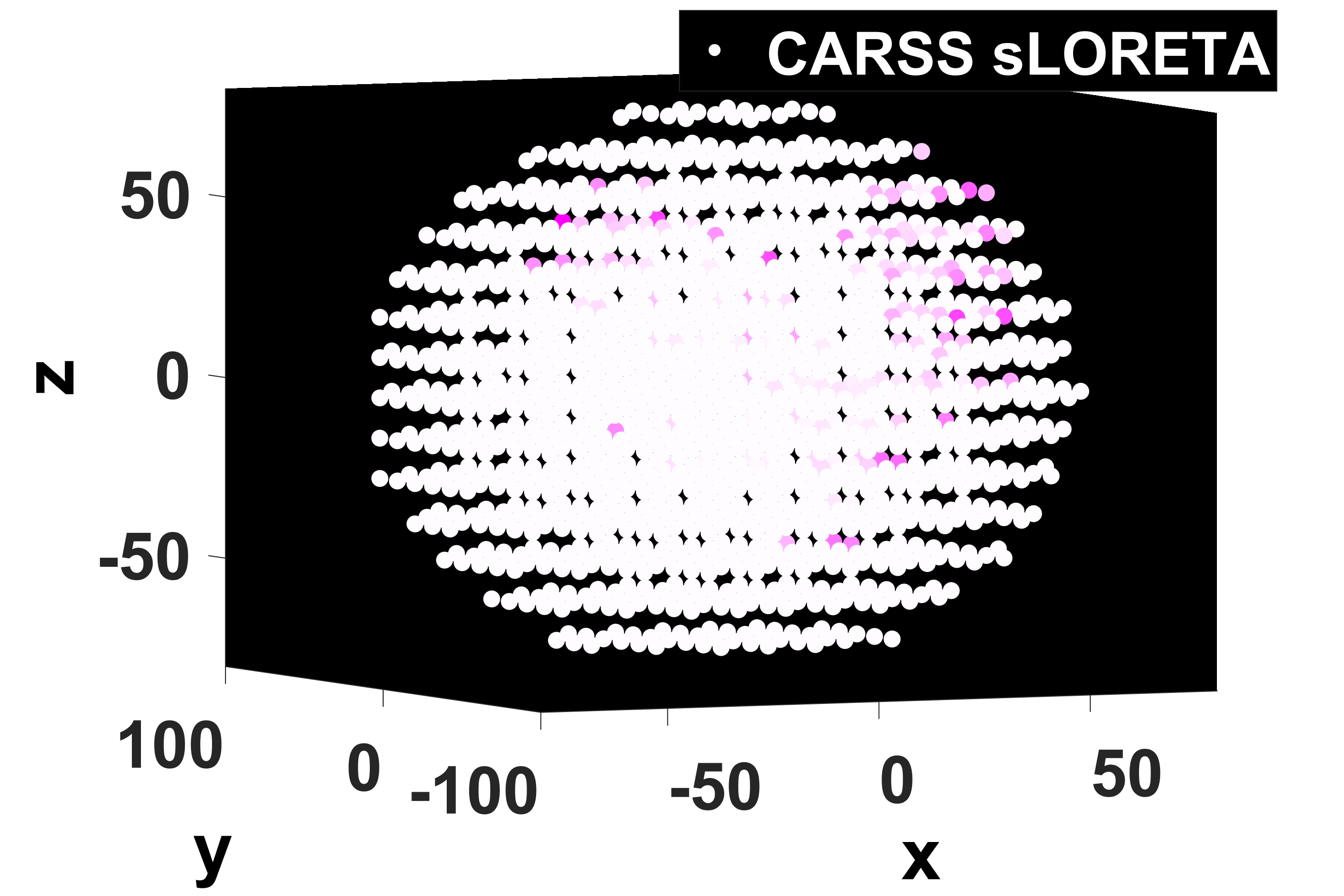}%
		\label{M46}}\hfill
	\caption{(a) The sources considered for test case-I (b)The prior most certain sources estimated in stage-I (c)Output of Stage-II with FOCUSS (d) FOCUSS without CARSS (e) sLORETA without CARSS (f) Output of stage-II with sLORETA for 17 dB noise}
	\label{M4}
\end{figure*}

\begin{figure*}[!t]
	\centering
	\subfloat{(a)}{%
		\includegraphics[width=3.33in]{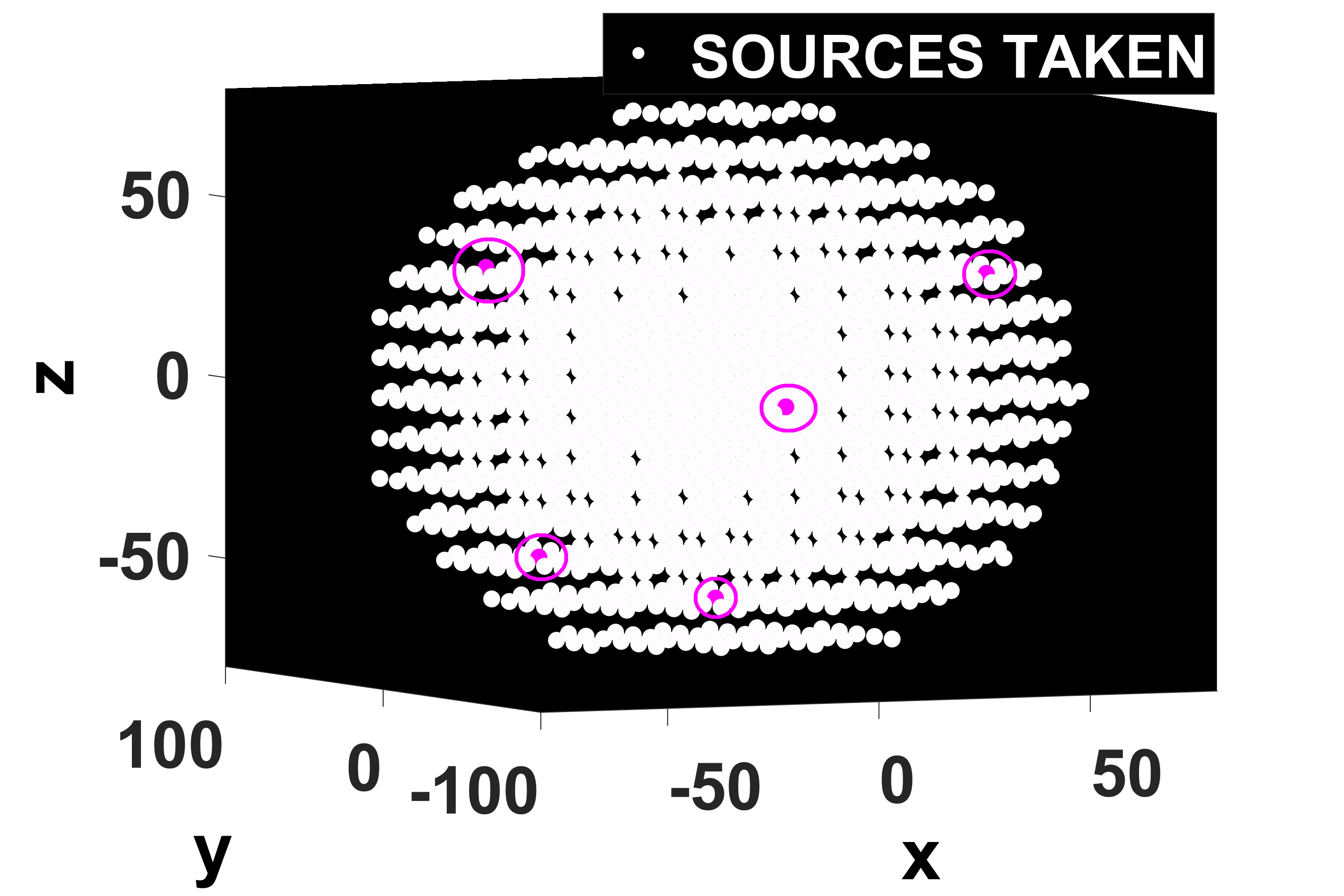}%
		\label{M51}}\hfill
	\subfloat{(b)}{%
		\includegraphics[width=3.33in]{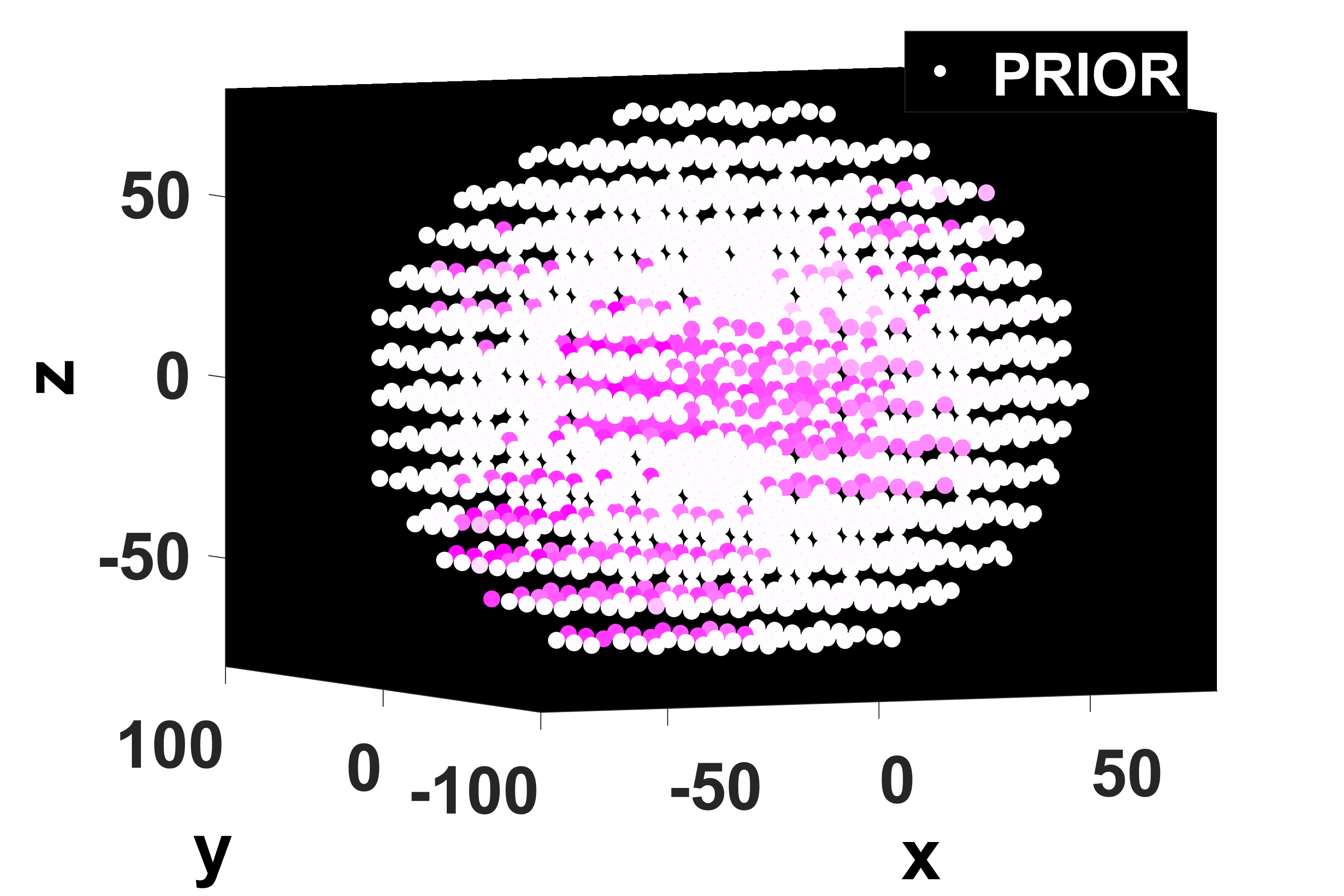}%
		\label{M52}}\hfill
	\subfloat{(c)}{%
		\includegraphics[width=3.33in]{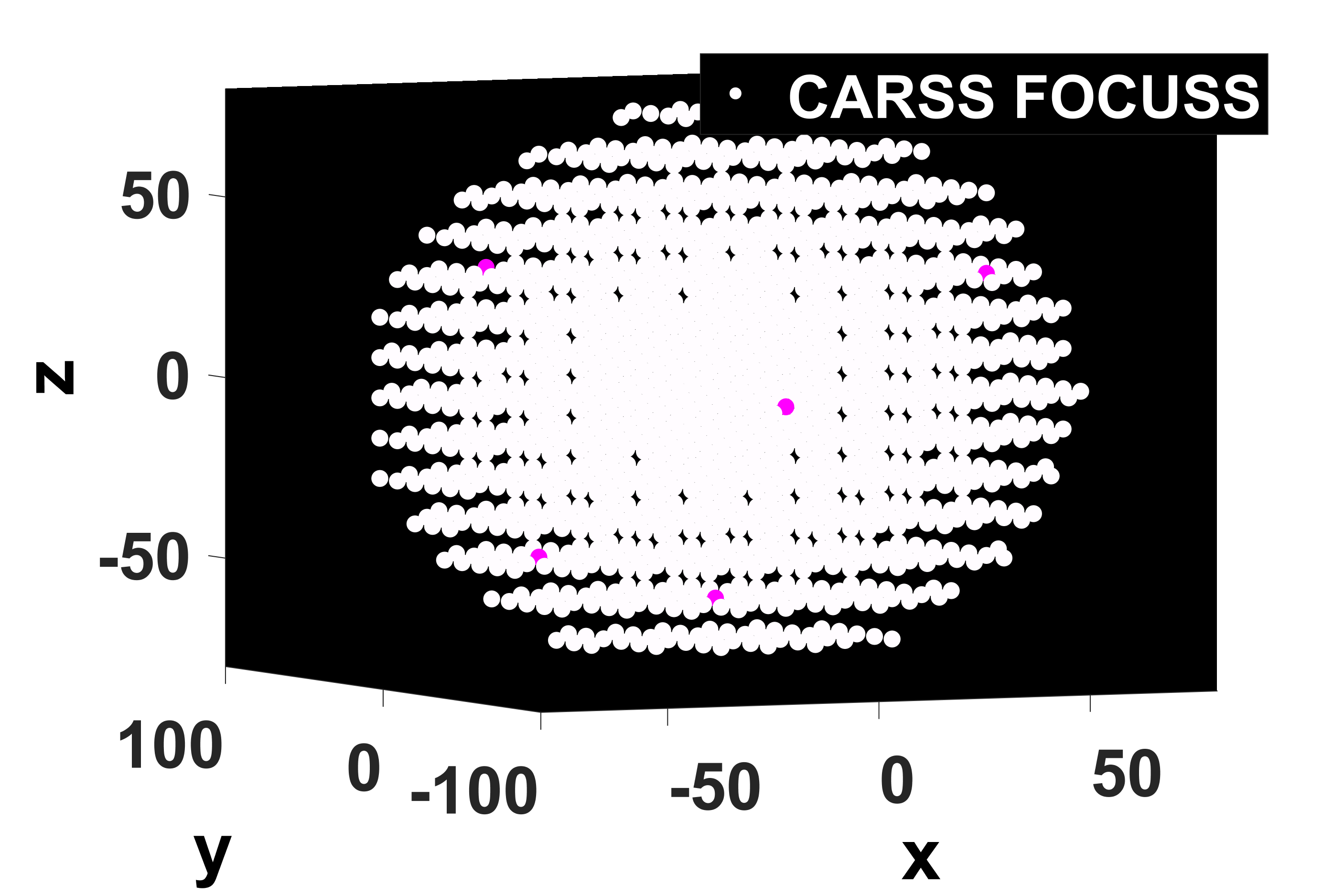}%
		\label{M53}}\hfill
	\centering
	\subfloat{(d)}{%
		\includegraphics[width=3.33in]{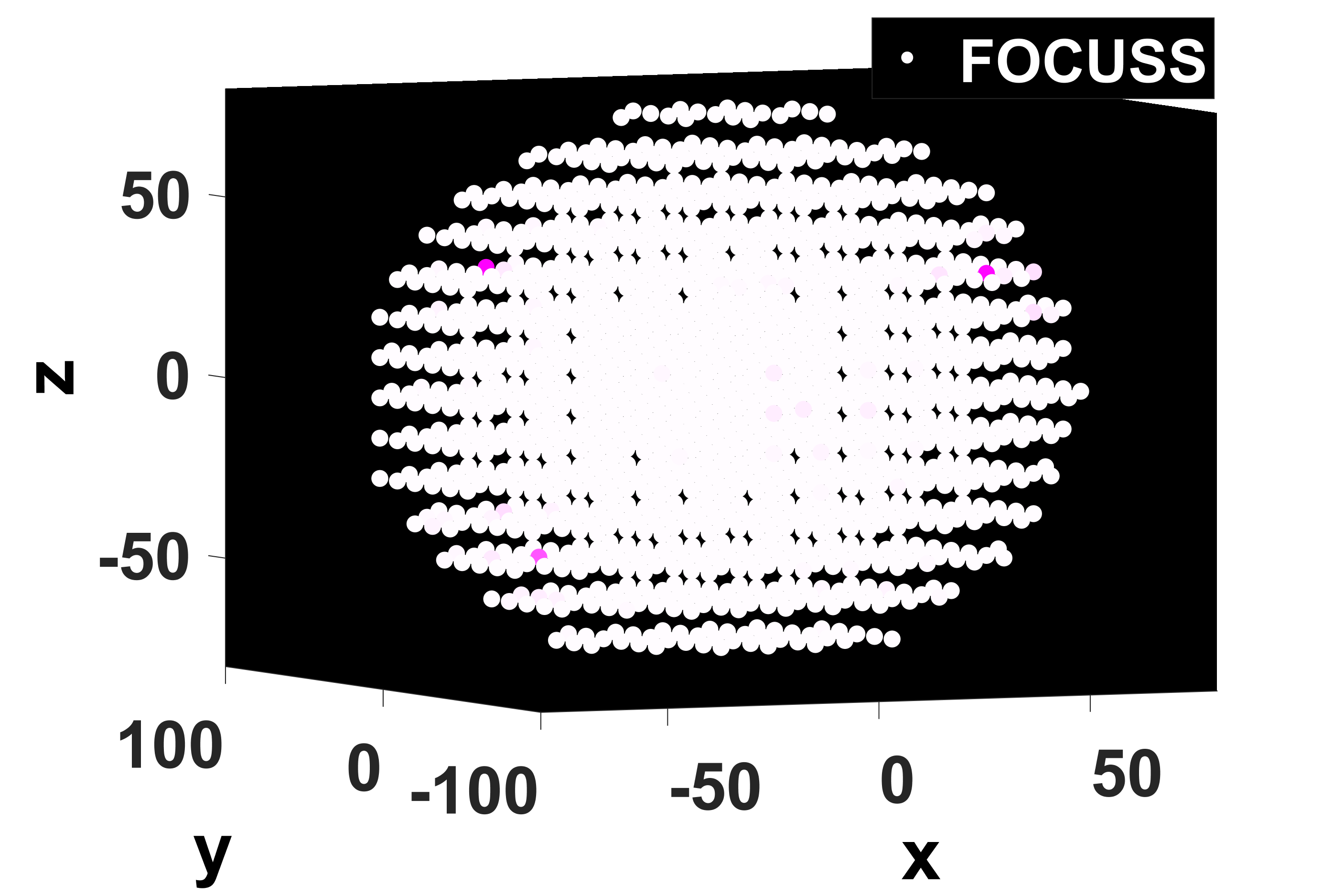}%
		\label{M54}}\hfill
	\subfloat{(e)}{%
		\includegraphics[width=3.33in]{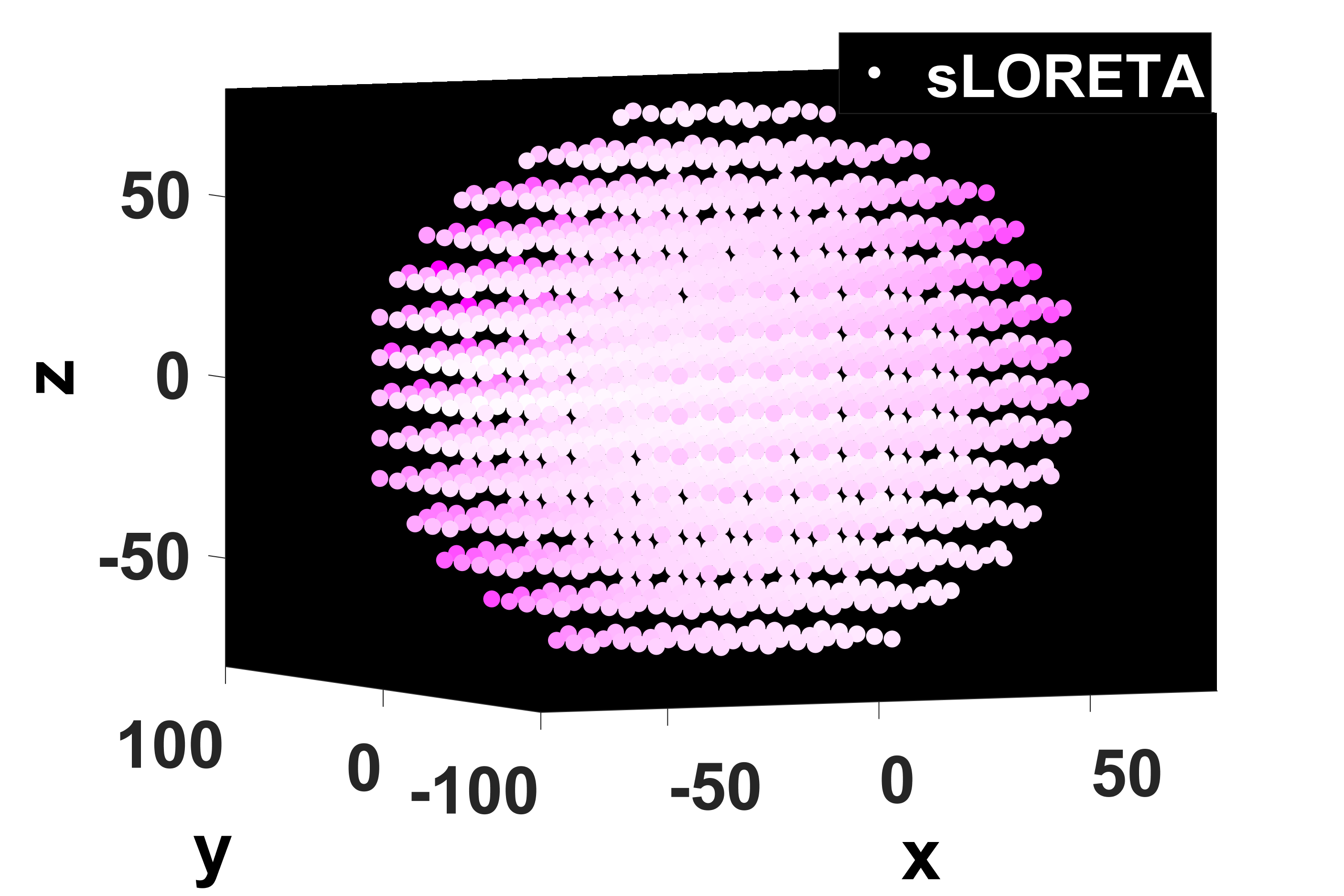}%
		\label{M55}}\hfill
	\subfloat{(f)}{%
		\includegraphics[width=3.33in]{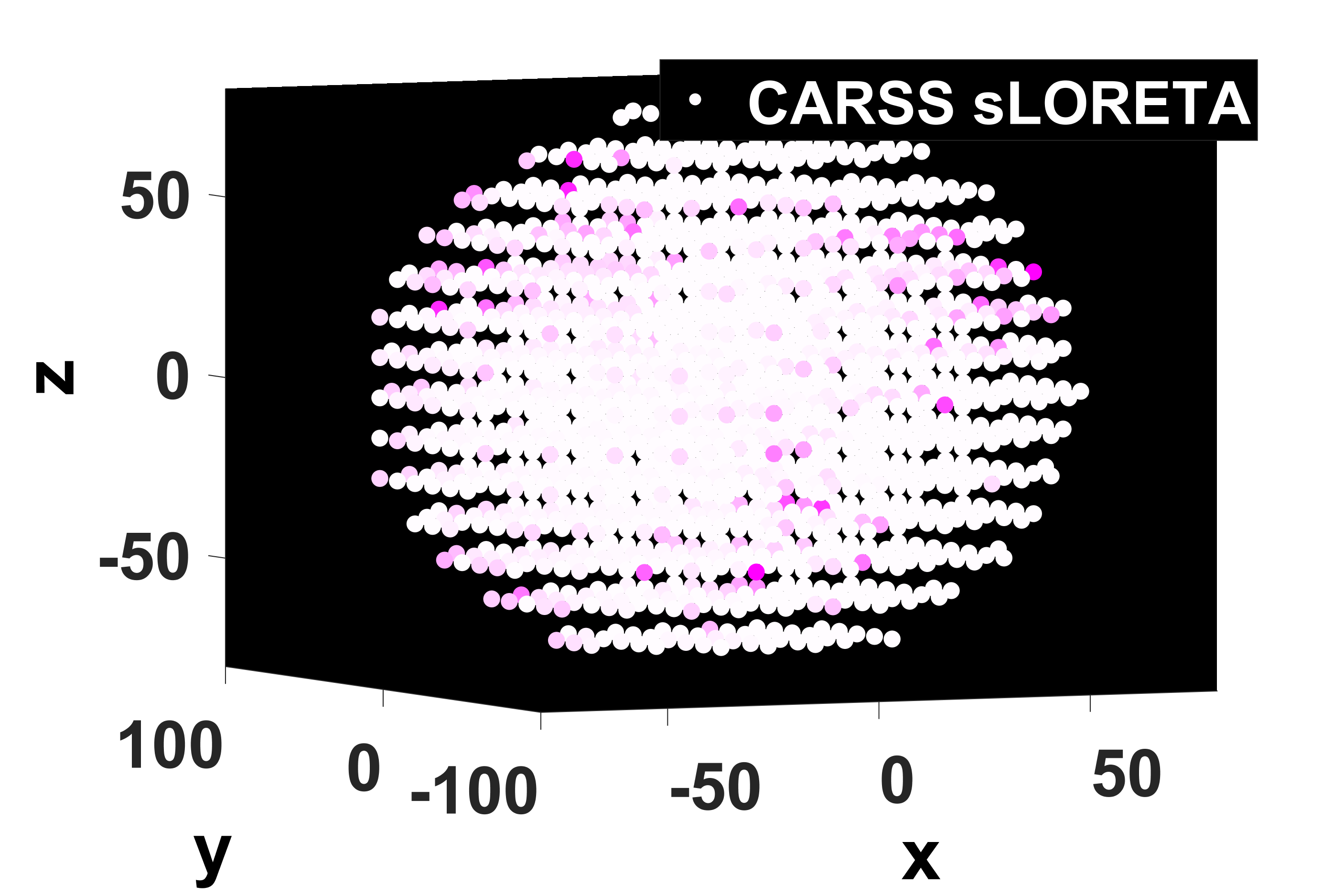}%
		\label{M56}}\hfill
	\caption{(a) The sources considered for Test Case-II are shown in magenta (all other sources in white) (b) Stage-I: The most certain sources (c)Output of Stage-II with FOCUSS (d)FOCUSS without CARSS (e)sLORETA without CARSS (f) Output of Stage-II with  sLORETA for 13 dB noise}
	\label{M5}
\end{figure*}

The $E_{localization}$, $E_{energy}$ and $NBI$ are chosen to evaluate the localization error, with what power is the source localized with and how much blurred the solution is. In all the figures, the pink dots shows the active/predicted sources with some moment and all sources i,e the solution space (or also can be predicted ones as inactive sources with zero moment) are shown in white dots. The active sources taken are shown as pink dots in pink circles (two in Fig. \ref{M41} and five Fig. \ref{M51}). $E.I$ refer to evaluation indexes in table \ref{TC1}, \ref{TCIII} and \ref{TCN}. The ideal evaluation indexes that have to be achieved are mentioned in the second column for tables \ref{TC1} and \ref{TCIII} whereas for table III, the same are mentioned under the first column in braces. 

In all the test cases presented, ideal value that has to be reached by $E_{localization}$ is $0$ $mm$ implying that any method with this has localized the source with zero error. The ideal value by $E_{energy}$ is $0$ implying that all energy with which the source is simulated is completely retrieved by that method. The ideal value to be achieved by $NBI$ is mentioned in the tables accordingly. An ideal value of $NBI$ implies that the source is localized finely without any blurring. 

\textbf{Test Case-I}: 6 active grid points are considered in 2 groups(focally extended sources). The group of three are near to each other with one grid-point distance. The source moments for all the sources are (1.1,1.2,1.3) along x,y and z directions. The sources chosen are from left surface top with coordinates [53.87 -24.22 31.56], [53.87-24.22 42.71] and [53.8671 -23.06 31.56] (labeled as S-I in table \ref{TC1}
and \ref{TCN}) and right surface bottom with coordinates [-13.06 65.02 31.56], [-13.06 65.02 42.71] and [-13.06 53.86 42.71](labeled as S-II). The evaluation indexes mentioned in Table \ref{TC1} and \ref{TCN} are the best ones of all the three in each group.  All distances mentioned in [] are in mm. 

In Stage-I, Eq.8 is used to find the $SI$. The solution space is reduced to 201 certain sources out of 4605 among which the active source(s) may lie (The pink ones in Fig. \ref{M42}). The lead field matrix $K$ of size $256\times 4605$ is reduced to $256\times 201$. These 201 sources are used in stage-II. 

The same test case is added with 17 dB SNR of random white noise. In Stage-I, for $SI$ Eq. 9 is used utilizing the robustness of  $l_2$ norm to noise. The solution space is reduced to 429 most certain sources out of which some might be active out of 4605 in total. The lead field matrix $K$ of size $256\times 4605$ is reduced to $256\times 429$. The 429 sources are only used in stage-II. 

The evaluation indexes (Fig. \ref{M4}, Table. \ref{TCN} and Table \ref{TC1}) show an improvement in accuracy of localizing the active sources for FOCUSS and sLORETA due to reduction of solution space (to most certain sources). 

\textbf{Test Case-II}: 5 active isolated grid points(3 deep and 2 surface) are considered. The sources are of more depth. Each source is excited with equal magnitude(5,5,5). The sources chosen are right surface top [53.86, -24.23, 31.56], right center deep [-1.9, -46.53,-1.9], left bottom deep [-35.37 20.40 -57.68], left surface top [-35.37 53.86 31.55], and center deep bottom [-1.91 -1.91 -57.68]. All the sources in Tables \ref{TC1} and \ref{TCN} are mentioned in the order given above as S-I, S-II, S-III, S-IV and S-V. FOCUSS missed out localizing S-II source, hence 'X' is mentioned in S-II column for FOCUSS in table \ref{TCIII}.  All distances mentioned in [] are in mm.

In Stage-I, for $SI$, Eq.8 is used. The solution space is reduced to 783 out of 4605 after the estimation of possibly certain sources using similarity index after Stage-I. This means there are only 783 most certain sources who might be active. The lead field matrix $K$ of size $256\times 4605$ is reduced to $256\times 783$. The 783 sources are only used in stage-II.

The same test case is added with 13 dB SNR of random white noise. In Stage-I, for $SI$ Eq. 9 is used using $l_2$ norm. The solution space is reduced to 1659 out of 4605 after the estimation of possibly certain sources using similarity index after Stage-I. This means there are only 1659 most certain sources who might be active. The lead field matrix $K$ of size $256\times 4605$ is reduced to $256\times 1659$. The 1659 sources are only used in stage-II.

The evaluation indexes for this test case are given in Table \ref{TCIII} (without noise added) and Table \ref{TCN} (with noise added) that shows an improvement in localization of the active sources for FOCUSS and sLORETA.  

\begin{table}[]
	\centering
	\caption{\textbf{Evaluation Indexes for Test-Case-I}}
	\label{TC1}
	\begin{tabular}{|c|c|c|c|c|c|}
		\hline
		\multirow{2}{*}{\textbf{E.I}} & 	& \multicolumn{2}{c|}{\textbf{S-I}} & \multicolumn{2}{c|}{\textbf{S-II}} \\ \cline{2-6}
		& \textbf{Act}	 & \textbf{FOCUSS} & {\textbf{C-FOCUSS}} & \textbf{FOCUSS} & {\textbf{C-FOCUSS}}     \\ \hline
		$\mathbf{E_{l}}$ &  0 & 0 & {0} & 0 & {0}    \\ \hline
		$\mathbf{E_{e}}$ & 0  & 0 & {0} & 0.1 & {0}       \\ \hline
		$\mathbf{NBI}$ & 2.6  & 3.7 &  {2.6}  & 4.0 & {2.6}     \\ \hline
	\end{tabular}
\end{table}


\begin{table*}[]
	\centering
	\caption{\textbf{Evaluation Indexes for Test-Case-II}}
	\label{TCIII}
	\begin{tabular}{|c|c|c|c|c|c|c|c|c|c|c|c|}
		\hline
		\multirow{2}{*}{\textbf{E.I}} &  &\multicolumn{2}{c|}{\textbf{S-I}} & \multicolumn{2}{c|}{\textbf{S-II}} & \multicolumn{2}{c|}{\textbf{S-III}} & \multicolumn{2}{c|}{\textbf{S-IV}} & \multicolumn{2}{c|}{\textbf{S-V}} \\ \cline{2-12}
		& \textbf{Act}	 & \textbf{FOCUSS} & \textbf{{C-FOCUSS}} & \textbf{FOCUSS} & \textbf{{C-FOCUSS}} & FOCUSS & \textbf{{C-FOCUSS}} & \textbf{FOCUSS} & \textbf{{C-FOCUSS}} & \textbf{FOCUSS} & \textbf{{C-FOCUSS}} \\ \hline
		$\mathbf{E_{l}}$ & 0  & 11.15  & 0 & X & 0 & 0 & 0 & 0 &  0 & 0 &  0  \\ \hline
		$\mathbf{E_{e}}$	& 0  & 0.6 & 0 & 1 & 0 & 0.3 & 0 & 0.4 &  0 & 0.9 &  0  \\ \hline
		$\mathbf{NBI}$& 4.4 & 8.6 & 4.4 & - & 4.4 & 5.2 & 4.4 & 6.0 & 4.4 & 73.4 & 4.4    \\ \hline
	\end{tabular}
\end{table*}

\begin{table*}[]
	\centering
	\caption{\textbf{Test Cases under Noise}}
	\label{TCN}
	\begin{tabular}{|c|c|c|c|c|c|c|c|c|c|c|c|c|c|c|}
		\hline
		\multirow{3}{*}{\textbf{E.I}} & \multicolumn{4}{c|}{\textbf{TC-I(17 dB)}} & \multicolumn{10}{c|}{\textbf{TC-II(13 dB)} } \\ \cline{2-15}
		&    \multicolumn{2}{c|}{\textbf{S-I}}    &   \multicolumn{2}{c|}{\textbf{S-II}}    &   \multicolumn{2}{c|}{\textbf{S-I}}    &   \multicolumn{2}{c|}{\textbf{S-II}}     &   \multicolumn{2}{c|}{\textbf{S-III}}  &  \multicolumn{2}{c|}{\textbf{S-IV}} &  \multicolumn{2}{c|}{\textbf{S-V}}  \\ \cline{2-15}
		& \textbf{C-sLOR} &\textbf{sLOR} &  \textbf{C-sLOR}  &\textbf{sLOR} &  \textbf{C-sLOR}    & \textbf{sLOR} &  \textbf{C-sLOR}   & \textbf{sLOR} & \textbf{C-sLOR}   &\textbf{sLOR} & \textbf{C-sLOR} &\textbf{sLOR} & \textbf{C-sLOR} & \textbf{sLOR}\\ \hline
		$\mathbf{E_{l}}$ (0) &   0  & 15.77 &   11.15   & 0 &  11.15    &  19.32 & 24.9    & 19.32 &  11.15  &  31.55  & 22.3 & 24.97 & 24.9 & 24.97\\ \hline
		$\mathbf{E_{e}}$ (0)& 0.6   & 0.9 &   0.4  & 0.9 &   0.66  & 0.9 &   0.9 & 0.9 & 0.75   &  0.9& 0.8 &  0.9& 0.5 & 0.9\\ \hline
		$\mathbf{NBI}$ &   0.6(0.8)   & 26.1 &  0.01     & 34.7 &  1.4(4.4) &  50.1 &    6.0  & 48.0 &  2.0 & 32.6 & 2.0 & 55.0 & 5.3 & 39.5 \\ \hline
	\end{tabular}
\end{table*}

\begin{figure*}[]
	\subfloat{}{%
		\includegraphics[width=2.35in]{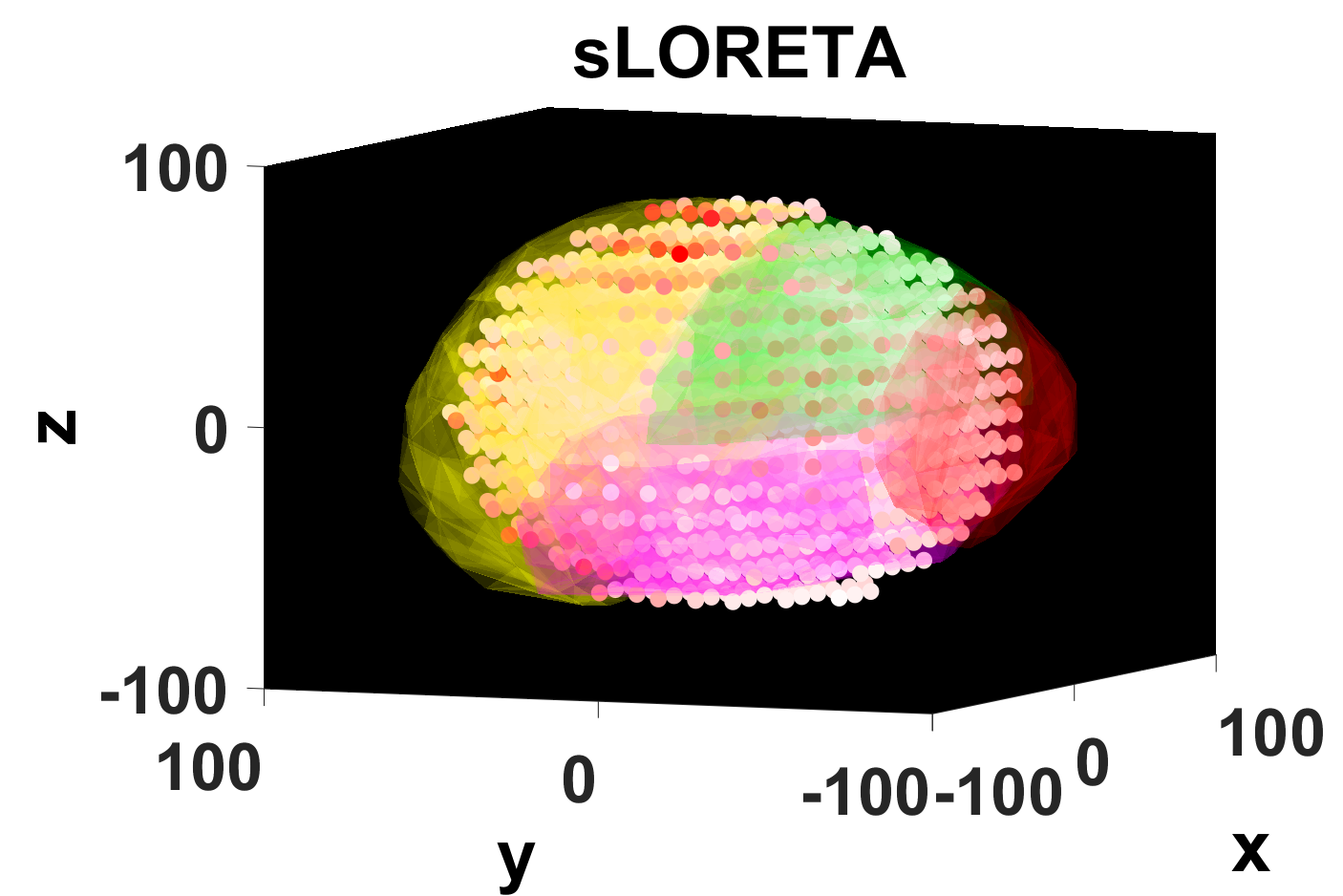}%
		\label{M61}}\hfill
	\subfloat{}{%
		\includegraphics[width=2.35in]{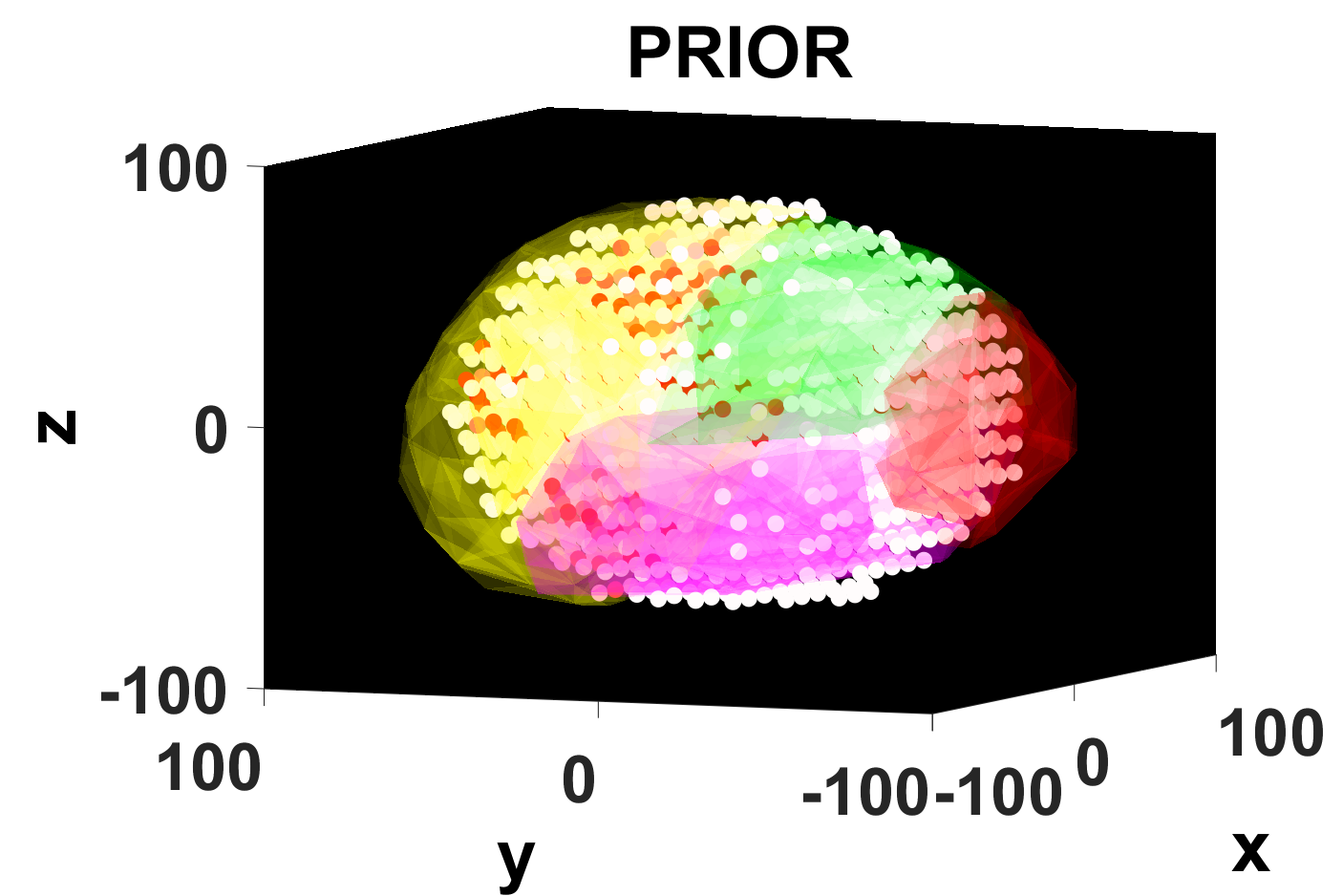}%
		\label{M62}}\hfill
	\subfloat{}{%
		\includegraphics[width=2.35in]{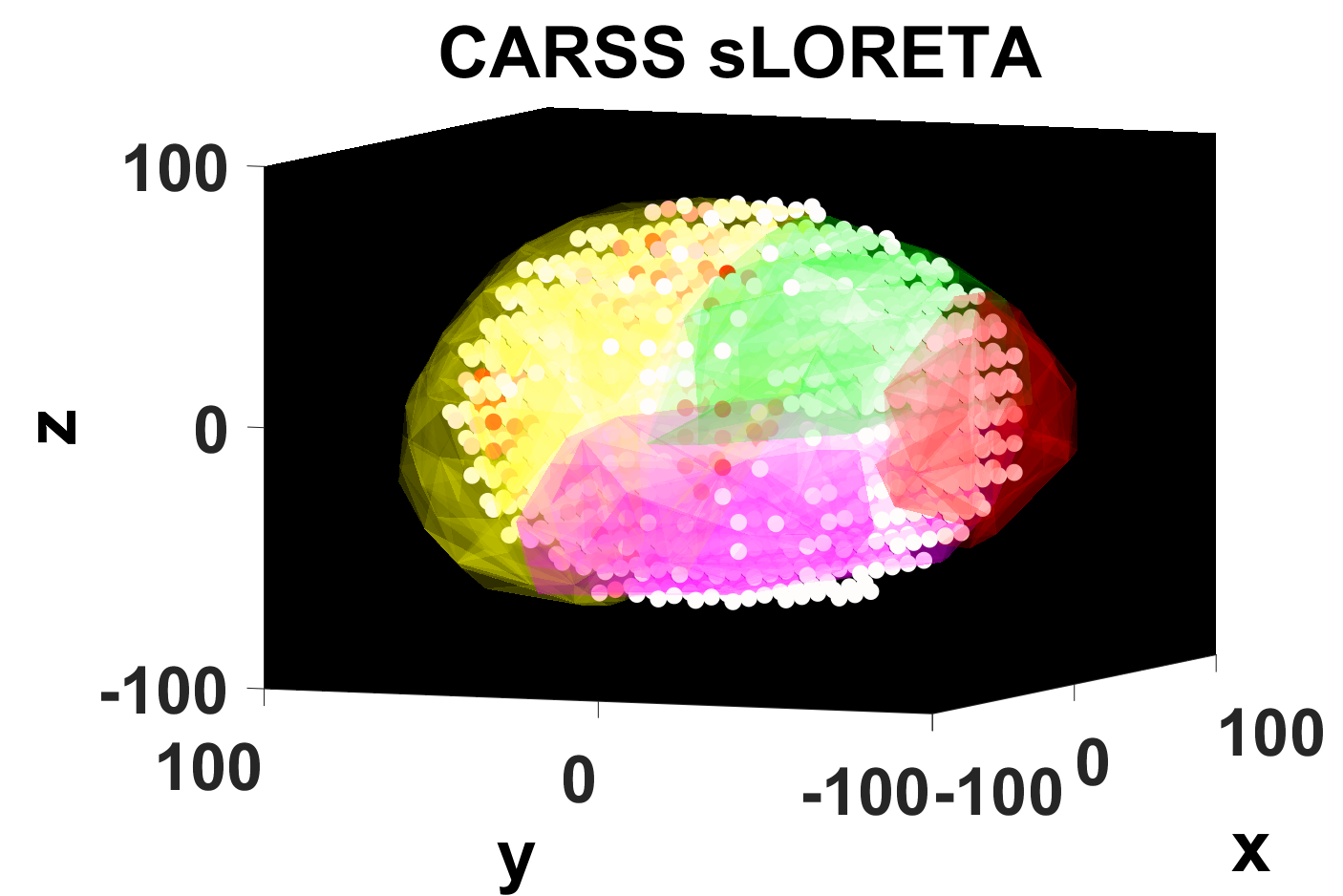}%
		\label{M63}}\hfill
	\caption{(a) sLORETA without using CARSS; The frontal, parietal, occipital and temporal are mentioned in yellow, green, pink and red. (b) Prior obtained when $10^{th}$ person is playing matching. (c) sLORETA is used in Stage-II for (b). }
	\label{M6}
\end{figure*}

\subsection{Real data tests}

The real data is obtained using 256 channel EGI hydrocel geodesic sensor nets EEG setup. There are total of 10 persons(subjects). Each person is made to play a cognitive game - Matching and EEG is recorded simultaneously. 

The ICA is performed using fieldtrip toolbox\cite{oostenveld2011fieldtrip}. The EOG artifacts are removed by finding $Pr's$ the ratio between power distributed on EEG channels prominently near to the eyes or prominent EOG artifact channels to EEG channels that are not selected as prominent EOG artifact channels using component to channel weight matrix generated by ICA. 

$Pr_i$ for all the components is calculated. It is defined as:
\begin{equation}
Pr_i = \frac{Pe_i-Pne_i}{Pne_i}
\end{equation}
where $Pe_i = \sum_{j=1}^{n}\mathbf{w_{i,j}}^2, j=1 \dots n$ and $Pne_i = \sum_{j=n}^{N}\mathbf{w_{i,j}}^2, j=n \dots N$. $j=1 \dots n$ are the EEG electrodes prominently near to the eyes. $Pr_i$ is power distribution ratio that gives the ratio of how much power is distributed at the probable EOG artifact electrodes to non-EOG electrodes for each component. $\mathbf{W}$ is the component to channel weight matrix generated by ICA. $Pe_i$ and $Pne_{i}$ are the power distributed at EOG and non-EOG channels. If $Pr \ge + \tau $, implies more power of the component in EOG electrodes making this a dominant EOG artifact. ECG artifacts are removed by observing the time course of the components. After this, ICA is projected back, filtered using a band pass filter of 4-30 Hz and followed by both temporal and spatial filtering.

The spatial filtering is done using the neighboring channels generated in stage-0.
\begin{equation}
\mathbf{\Phi_i^s(t)} = \frac{1}{L}\sum_{j=1}^{L}\mathbf{\Phi_i(t)}
\end{equation}
where $L$ is the number of neighbors of $i^{th}$ channel and $\mathbf{\Phi_i^s(t)}$ is the measurement vector after spatial filtering. Temporal filtering is applying moving average filter along the time axis.

The cortical activity predicted for $10^{th}$ person are presented in Fig. \ref{M6}. The prior (in Fig. \ref{M6}(a)) corresponding to the cognitive game(matching) shows profound activity in the frontal lobe which is associated with logical thinking of a person\cite{nolte2002human}. These observations are also confirmed by sLORETA in Fig. \ref{M6}(c). The number of certain sources predicted by the prior is 662 out of 4605.

\section{Conclusions}

As the results show, two goals of this method are achieved. Firstly, reduction of solution space to most certain sources that might be active and hence eliminating the redundancy while solving the problem. Secondly, localization accuracy of other methods is improved thereby achieving the main goal of identifying the active sources in the brain accurately. 


\bibliographystyle{IEEEtran}	
\bibliography{IEEEabrv,ref}

\end{document}